\documentclass{iopart}
\usepackage{iopams}
\usepackage{graphicx}
\usepackage{afterpage}

\newcommand{\uvozovky}[1]{``#1''}

\def\beq{\begin{equation}}
\def\eeq{\end{equation}}
\def\bea{\begin{eqnarray}}
\def\eea{\end{eqnarray}}

\let\phi=\varphi

\let\phi=\varphi
\let\rho=\varrho

\begin{document}

\title[Non-monotonic Keplerian velocity profiles around brany Kerr black holes]{Non-monotonic Keplerian velocity profiles around near-extreme braneworld Kerr black holes}

\author{ Zden\v{e}k Stuchl\'{\i}k, Martin Blaschke and Petr Slan\'{y} }

\address{Institute of Physics, 
         Faculty of Philosophy and Science, 
         Silesian university in Opava, 
         Bezru\v{c}ovo n\'{a}m. 13, CZ-746 01 Opava, Czech Republic}
           
\ead{MartinBlaschke@seznam.cz, Petr.Slany@fpf.slu.cz, Zdenek.Stuchlik@fpf.slu.cz}

\begin{abstract}
We study the~non-monotonic Keplerian velocity profiles related to locally non-rotating frames (LNRF) in~the~field of~near-extreme braneworld Kerr black holes and naked singularities in~which the~non-local gravitational effects of~the~bulk are represented by a~braneworld tidal charge $b$ and the~4D geometry of~the~spacetime structure is governed by the~Kerr--Newman geometry. We show that positive tidal charge has a~tendency to~restrict the~values of~the~black hole dimensionless spin $a$ admitting existence of~the~non-monotonic Keplerian LNRF-velocity profiles; the~non-monotonic profiles exist in~the~black hole spacetimes with tidal charge smaller than $b=0.41005$ (and spin larger than $a=0.76808$). With decreasing value of~the~tidal charge (which need not be only positive), both the~region of spin allowing the~non-monotonicity in~the~LNRF-velocity profile around braneworld Kerr black hole and the~velocity difference in~the~minimum-maximum parts of the~velocity profile increase implying growing astrophysical relevance of this phenomenon.  
\end{abstract}

\pacs{04.70.-s, 04.20.-q, 98.80.-k, 04.50.-h}

\submitto{\CQG}

\maketitle

\section{Introduction}

Fast rotating black holes play a~crucial role in understanding processes observed in quasars and Active Galactic Nuclei (AGN) or in~microquasars. It has been shown that supermassive black holes in AGN evolve into states with dimensionless spin $a \sim 1$ due to accretion from thin discs \cite{Vol-Mad-Qua-Ree:2005:apj:, Sha:2005:apj:}. This statement is supported by analysis of profiled X-ray (Fe56) lines observed in some AGN (e.g. in MCG-6-30-15) \cite{Tan-etal:1995:NATURE:, Miy-atal:2009:arxiv:,Rey-etal:2009:} and in some microquasars (e.g., GRS 1915+105) \cite{McCli-etal:2006:astro-ph/0606076:}. Evidence for the~existence of near-extreme Kerr black holes comes from high-frequency quasi-periodic oscillations (QPOs) of observed X-ray flux in some microquasars \cite{Tor-Abr-Klu-Stu:2005:, Ste-Mcc-etal:2008:}. A~fast rotating black hole could be also located in the~Galaxy center source Sqr $\mathrm{A^*}$ \cite{Asch:2004:ASTRA:,Tor:2005:ASTRN:,Mey-etal:2006:ASTRA:}. 

It is widely accepted that the~phenomena observed in AGN and microquasars are related to accretion discs orbiting Kerr black holes. However, we can consider also the~possibility to explain these phenomena by Kerr superspinars with external field described by the~geometry of Kerr naked singularity spacetime \cite{Gim-Hor:2009:Phys.LettB:}. Then both accretion and related optical effects and the~QPOs effects enable us to find clear signature of the~Kerr superspinar presence \cite{DeF:1974:aap:,DeF:1978:Nat:,Stu:1980:BULAI:,Stu:1981:BULAI:,Stu-Sche:2010:CLAQG:,Stu-Hle-Tru:2011:CLAQG:}.

\begin{figure}[ht]
\begin{center}
\begin{minipage}{.49\linewidth}
\centering
\includegraphics[width=\linewidth]{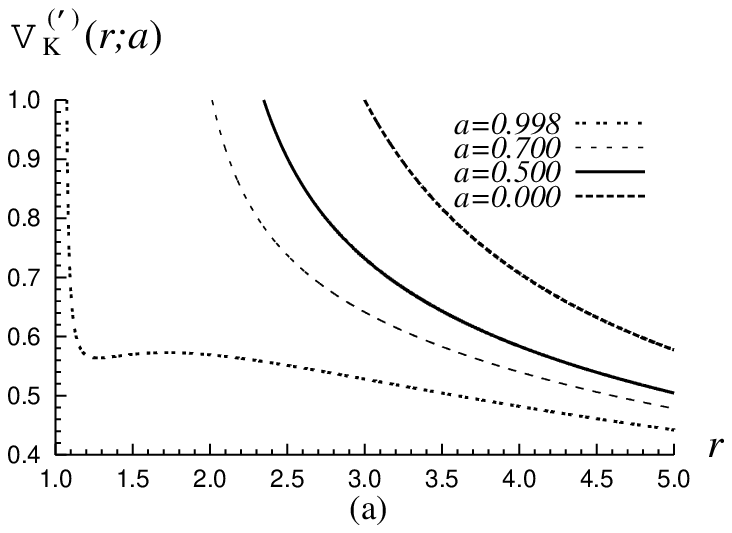}
\end{minipage}\hfill
\begin{minipage}{.49\linewidth}
\centering
\includegraphics[width=\linewidth]{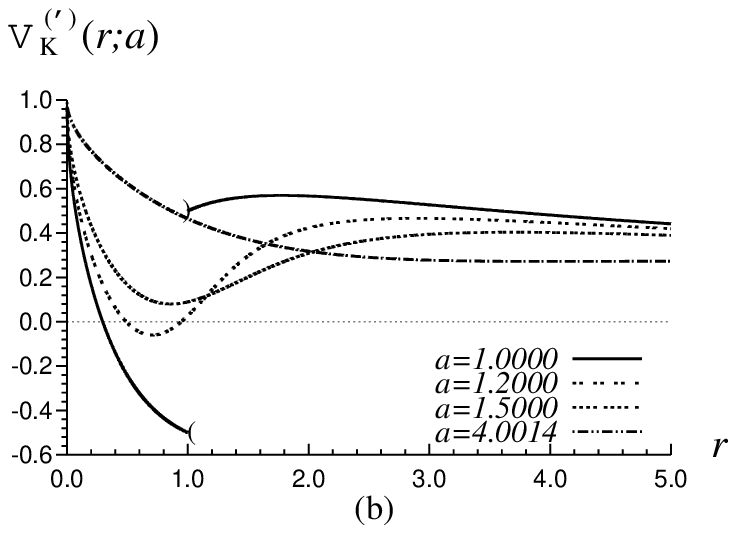}
\end{minipage}
\caption{\label{class}Keplerian velocity profiles related to the~LNRF. 

(a): Kerr black holes -- the~velocity profiles presented for some values of the~black hole spin. The~Aschenbach effect appears for near-extreme black holes and is weak. 

(b): Kerr naked singularities -- the~velocity profiles are given for some values of the~spin, demonstrating existence of Aschenbach effect for orbits with negative valued velocity. For completeness, the~velocity profile is given also for extreme black hole, demonstrating velocity jump at $r=1$.}
\end{center}
\end{figure}

Properties of accretion discs can be appropriately represented by circular orbits of test particles or fluid elements orbiting black holes (superspinars). The~local properties can be efficiently expressed when related to the~locally non-rotating frames (LNRF), since these frames corotate with the~spacetime in a~way that enables to cancel the~frame-dragging effects as much as possible \cite{Bar-Pre-Teu:1972:ASTRJ2:}. A~new phenomenon related to the~LNRF-velocity profiles of matter orbiting near-extreme Kerr black holes has been found by B. Aschenbach \cite{Asch:2004:ASTRA:,Asch:2008:ChinJP:,Stu-Sla-Tor-Abr:2005:PRD:}, namely a~non-monotonicity in the~velocity profile of the~Keplerian motion in the~field of Kerr black holes with dimensionless spin $a>0.9953$. Such a~hump in the~LNRF-velocity profile of the~corotating orbits is a~typical and relatively strong feature in the~case of Keplerian motion in the~field of Kerr naked singularities, but in the~case of Kerr black holes it is a~very small effect appearing for near-extreme black holes only -- see Figure \ref{class}. In the~naked singularity case we call the~orbits to be of 1st family rather than corotating, since these can be retrograde relative to the~LNRF in vicinity of the~ring singularity for small values of spin ($a < 5/3$), while they are corotating for larger values of spin \cite{Stu:1980:BULAI:}; the~humpy character of the~LNRF-velocity profile ceases for naked singularities with $a > 4.0014$ -- as demonstrated in the~Figure \ref{class}. 
A~study of non-Keplerian distribution of specific angular momentum ($l$ = const), related to geometrically thick discs of perfect fluid, has shown that the~\uvozovky{humpy} LNRF-velocity profile appears for near-extreme Kerr  black holes with $a>0.9998$ \cite{Stu-Sla-Tor-Abr:2005:PRD:}. The~humpy LNRF-velocity profile emerges in the~ergosphere of near-extreme Kerr black holes, at vicinity of the~marginally stable circular orbit. Maximal velocity difference between the~local minimum and maximum of the~humpy Keplerian velocity profiles is $\Delta v \approx 0.07\,c$ and takes place for  $a=1\,$ \cite{Stu-Sla-Tor:2007:ASTRA:}.

Using idea of \uvozovky{hump-induced} oscillations related to humpy LNRF-velocity profiles, the~Extended Orbital Resonance Model (EXORM) was developed and applied to explain complex QPO patterns observed in some black hole sources \cite{Stu-Sla-Tor:2007:ASTRA:,Stu-Sla-Tor:2007a:ASTRA:,Sla-Stu:2008:ASTRA:}. In the~EXORM, the~resonant phenomena between oscillations with orbital frequencies (Keplerian and epicyclic -- radial or vertical) and so called \uvozovky{humpy} frequency, given by the~maximal slope of the~LNRF-velocity profile in its humpy part, are assumed to appear at the~radius where the~\uvozovky{humpy} oscillations are expected to be generated by the~non-monotonicity of the~LNRF-velocity profile \cite{Stu-Sla-Tor:2007a:ASTRA:}. The~EXORM is able to explain all five high-frequency QPOs observed in the~microquasar GRS 1915+105 by the~humpy frequency, the~radial (and vertical) epicyclic frequencies and their simple combinations taken at the~common \uvozovky{humpy} radius, implying the~black hole parameters of the~source $M=14.8\,M_\odot $, a=0.9998 \cite{Stu-Sla-Tor:2007:ASTRA:}, in good agreement with estimates given by different methods \cite{McCli-etal:2006:astro-ph/0606076:,Asch:2008:ChinJP:}. This model can give interesting results also in the~case of the~X-ray binary system XTE J1650-500 \cite{Sla-Stu:2008:ASTRA:}, and an~ULX candidate system NGC 5408 X-1. On the~other hand, QPOs observed in Sqr $\mathrm{A^*}$ \cite{Asch:2004:ASTRA:} cannot be explained by the~EXORM \cite{Stu-Sla-Tor:2008:MG11:CrossRef}.     

The~humpy LNRF-velocity profiles for both Keplerian and $l=\mathrm{const}$ specific angular momentum distributions of orbiting matter were studied in the~Kerr-de Sitter and Kerr-anti-de Sitter black hole spacetimes and values of the~black hole spin $a$ allowing for existence of the~humpy profiles were found in dependence on the~value of the~cosmological constant \cite{Mul-Asch:2007:CLAQG:NonMonoVel,Sla-Stu:2007:CLAQG:}. 

Last decade gave rise to a~plenty of models modifying the~4D Einstein general relativity due to hidden dimensions, therefore, it is interesting to investigate the~Aschenbach effect in rotating black hole (naked singularity) spacetimes  allowed in alternative gravitational theories. The~string theories, describing gravity as higher-dimensional interaction appearing to be effectively 4D at low energies, inspired braneworld models assuming the~observable universe to be a~3-brane, i.e. \uvozovky{domain wall}, to which the~non-gravitational matter fields are confined, while gravity enters the~extra spatial dimensions that could be much larger than 
$l_p\sim 10^{-33}\,$cm. The~model of Randall and Sundrum (RS model) \cite{Ran-Sun:1999:PHYSR4:LarMassHie} allows gravity localized near the~brane with an~infinite size extra dimension while the~warped spacetime satisfies the~5D Einstein equations with negative cosmological constant. An~arbitrary energy-momentum tensor could then be allowed on the~brane and effective 4D Einstein equations have to be satisfied on the~brane. The~RS model implies standard 4D Einstein equations in the~low energy limit, but significant deviations occur at high energies, near black holes or compact stars. The~combination of high-energy (local) and bulk stress (non-local) effects alters the~matching problem on the~brane in comparison with the~standard 4D gravity \cite{Ger-Maa:2001:PRD:}. The~bulk gravity stresses imply that the~matching conditions do not have unique solution on the~brane and the~5D Weyl tensor is needed as a~minimum condition for uniqueness.

No exact solution of the~5D braneworld Einstein equations is known at present, but a~numerical solution has been found quite recently \cite{Fig-Wis:2011:Arxiv:}. On the~other hand, 4D stationary and axisymmetric vacuum solution describing a~braneworld rotating black hole has been found by solving the~braneworld  constrained equations under an~assumption of specialized form of the~metric (namely of the~Kerr-Schild form) \cite{Ali-Gum:2005:}. Of course, it is not an~exact solution satisfying the~full system of 5D equations, but in the~framework of the~constrained equations it represents a~consistent rotating black hole solution reflecting the~influence of the~extra dimension through a~single braneworld parameter. The~braneworld rotating black holes are described by the~metric tensor of the~Kerr--Newman form with the~braneworld tidal charge $b$ determining the~5D non-local gravitational coupling between the~brane and the~bulk \cite{Ali-Gum:2005:}. For non-rotating braneworld black holes, the~metric is reduced to the~Reissner-Nordstr$\mathrm{\ddot o}$m form containing the~tidal charge \cite{Dad-etal:2002:PHYLB:}. This spacetime can also represent the~external field of braneworld neutron stars described by the~uniform density internal spacetime \cite{Ger-Maa:2001:PRD:}. Influence of the~braneworld tidal charge on physical processes has been extensively investigated for both the~black holes \cite{Stu-Kot:2009:GRG:,Ali-Tal:2009:PRD:,Sche-Stu:2009:GRG:,Sche-Stu:2009:IJMPD:,Abd-Ahm:2010:PRD:,Bin-Nun:2010:PHYSR4:} and neutron stars \cite{Kot-Stu-Tor:2008:CLAQG:,Mam-Hak-Toj:2010:MPLA:,Mor-Ahm-Abd-Mam:2010:ASS:,Mor-Ahm:2010:ArXiv:,Hla-Stu:2011:JCAP:}, or in the~weak field limit \cite{Boh-etal:2008:CLAQG:,Boh-DeR:2010:arxiv:}. In the~case of microscopic black holes, an~experimental evidence is assumed in LHC \cite{CMS:2010:Arxiv:}. 

The~braneworld tidal charge can be, in principle, both positive and negative, but the~negative values are probably more relevant \cite{Dad-etal:2002:PHYLB:}. Notice that for $b>0$ the~braneworld spacetime can be identified with the~Kerr--Newman spacetime by $b \rightarrow Q^2\,$, where $Q^2$ is the~squared electric charge; however it is not the~Kerr--Newman background since the~electromagnetic part of this background is missing. Some astrophysically relevant restrictions on the value of the~tidal charge $b$ were obtained both in the~weak-field limit \cite{Boh-etal:2008:CLAQG:} and in the strong-field limit  \cite{Kot-Stu-Tor:2008:CLAQG:}. 

Here, we shall study existence of the~humpy LNRF-velocity profiles in the~field of braneworld rotating black holes considering both negative and positive values of the~braneworld tidal charge. Our results related to $b>0$ are relevant also in the~case of the~standard Kerr--Newman spacetimes (with $b \rightarrow Q^2\,$), for uncharged particles. We restrict our attention to the~Keplerian LNRF-velocity profiles postponing the~study of perfect fluid configurations to future work.

\section{Effective gravitational equations in braneworld models}

In the~5D warped space models of Randall and Sundrum, the~gravitational field equations in the~bulk can be expressed in the~form \cite{Dad-etal:2002:PHYLB:,Shi-Mae-Sas:2000:PHYSR4}

\begin{equation}
 \tilde{G}_{AB}=\tilde{k}^2[-\tilde\Lambda \tilde g_{AB}+\delta(\chi)(-\lambda g_{AB}+T_{AB})],\label{beq1}
\end{equation}
where the~fundamental 5D Planck mass $\tilde M_\mathrm{P}$ enters via $\tilde{k}^2=8\pi/\tilde{M}_\mathrm{P}^3 $, $\lambda$ is the~brane tension, and $\tilde\Lambda$ is the~negative bulk cosmological constant; $g_{AB}=\tilde{g}_{AB}-n_A n_B$ is the~induced metric on the~brane, with $n_A$ being the~unit vector normal to the~brane.
\par
The~effective gravitational field equations induced on the~brane are determined by the~bulk field equations (\ref{beq1}), the~Gauss--Codazzi equations and the~generalised matching Israel conditions. They can be expressed in the~form of modified  Einstein's equations containing additional terms reflecting bulk effects onto the~brane \cite{Dad-etal:2002:PHYLB:,Shi-Mae-Sas:2000:PHYSR4}

\begin{figure}[ht]
\includegraphics[width=\linewidth]{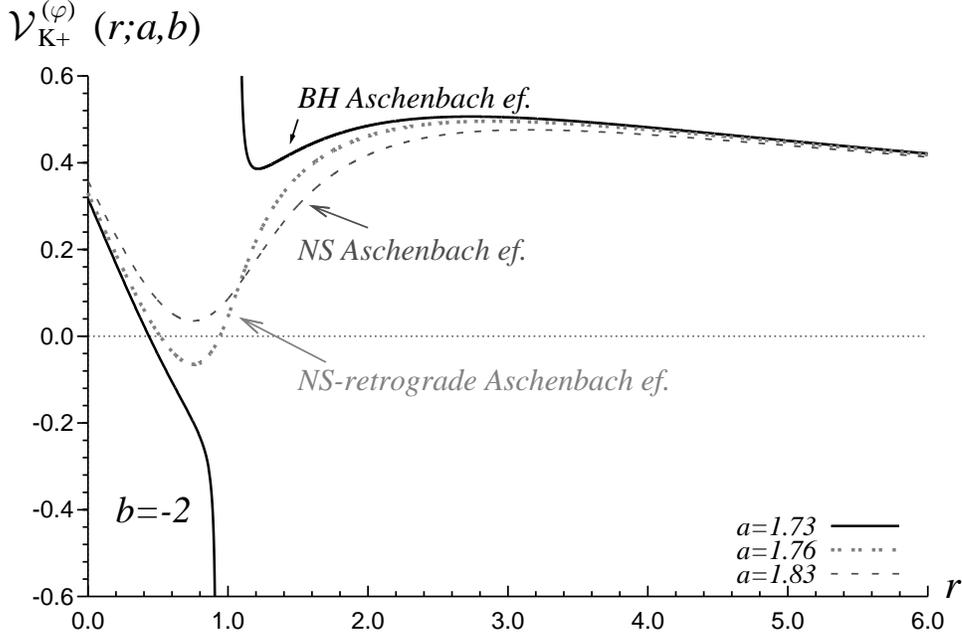}
\caption{\label{Ovkns} Non-monotonic plus-family orbital velocity profiles in braneworld Kerr black hole and naked singularity spacetimes with
the~tidal charge $b=-2$, given for three appropriately chosen values of spin $a$ reflecting the~whole variety of possible behaviour related to the~Aschenbach effect.}
\end{figure}

\begin{figure}[t]
\begin{center}
\includegraphics[width=0.9\linewidth]{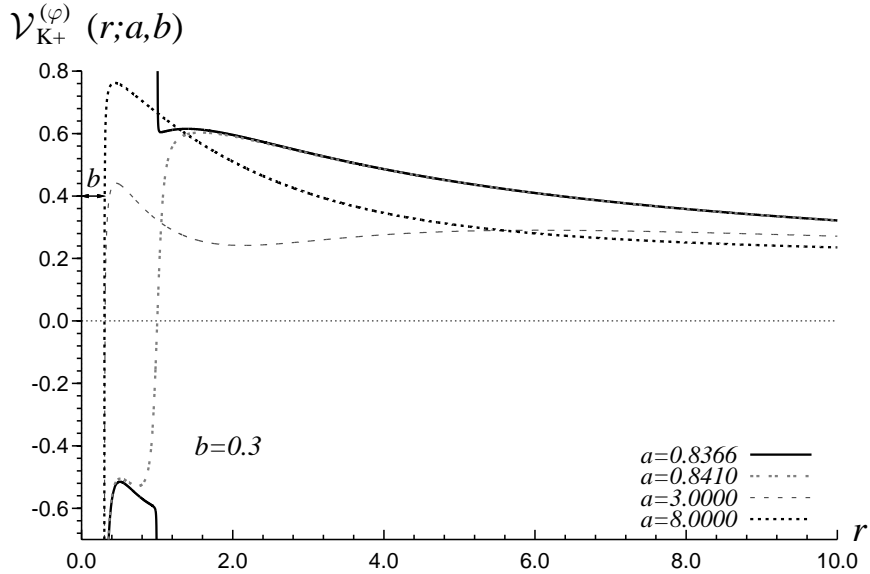}
\caption{\label{Vkod} Non-monotonic plus-family orbital velocity profiles in braneworld Kerr black hole and naked singularity spacetimes with 
the~tidal charge $b=0.3$.} 
\end{center}
\end{figure}

\begin{equation}\label{beq2}
 G_{\mu\nu}=-\Lambda g_{\mu\nu}+k^2 T_{\mu\nu} + \tilde{k}^2 S_{\mu\nu} -\mathcal{E}_{\mu\nu}\, ,
\end{equation}
where  $k^2=8\pi/M_\mathrm{P}^2$, with $M_\mathrm{P}$ being the~braneworld Planck mass. The~relations of the~energy scales and cosmological constants are given in the~form

\begin{equation}\label{beq3}
 M_\mathrm{P}=\sqrt{\frac{3}{4\pi}}\left(\frac{\tilde{M}_\mathrm{P}^2}{\sqrt{\lambda}}\right)\tilde{M}_\mathrm{P};\quad \Lambda=\frac{4\pi}{\tilde{M}_\mathrm{P}^3}\left[\tilde\Lambda+\left(\frac{4\pi}{3\tilde{M}_\mathrm{P}^3}\right)\lambda^2\right], .
\end{equation}
Local bulk effects on the~matter are determined by the~``squared energy-momentum'' tensor $S_{\mu\nu}$, 
while the~non-local bulk effects are given by the~tensor $\mathcal{E}_{\mu\nu}$. 

\par
Assuming zero cosmological constant on the~brane ($\Lambda=0$) we arrive to the~condition

\begin{equation}\label{beq5}
 \tilde\Lambda=-\frac{4\pi\lambda^2}{3\tilde{M}_\mathrm{P}^3}\, .
\end{equation}
In the~vacuum case, $T_{\mu\nu}=0=S_{\mu\nu}$, the~effective gravitational field equations on the~brane reduce to the~form \cite{Shi-Mae-Sas:2000:PHYSR4}
\begin{equation}
 R_{\mu\nu}=-\mathcal{E}_{\mu\nu}\, ,\quad R_\mu^{\phantom{\mu}\mu}=0=\mathcal{E}_\mu^{\phantom{\mu}\mu}\label{beq6}
\end{equation}
implying divergence constraint \cite{Shi-Mae-Sas:2000:PHYSR4}

\begin{equation}
 \nabla^\mu\mathcal{E}_{\mu\nu}=0\label{beq7}
\end{equation}
where $\nabla_{\mu}$ denotes the~covariant derivative on the~brane.
\par
The~equation (\ref{beq7}) represents Bianchi identities on the~brane, i.e., an~integrability condition for the~field equations $R_{\mu\nu}=-\mathcal{E}_{\mu\nu}$ \cite{Ali-Gum:2005:}. For stationary and axisymmetric (or static, spherically symmetric) solutions, (\ref{beq6}) and (\ref{beq7}) form a~closed system of equations on the~brane. 
\par
The~4D general relativity energy-momentum tensor $T_{\mu\nu}$ (with $T_\mu^{\phantom{\mu}\mu}=0$) can be formally identified to the~bulk Weyl term on the~brane due to the~correspondence 

\begin{equation}
 k^2 T_{\mu\nu}\quad\leftrightarrow\quad -\mathcal{E}_{\mu\nu}\, .\label{beq8}
\end{equation}
The~general relativity conservation law $\nabla^\mu T_{\mu\nu}=0$ then corresponds to the~ constraint equation on the~brane (\ref{beq7}). This behaviour indicates that the~Einstein-Maxwell solutions in standard general relativity should correspond to constrained braneworld vacuum solutions. This was indeed shown in the~case of braneworld (Reissner--Nordstr\"{o}m and Kerr--Newman) black hole solutions \cite{Ali-Gum:2005:}. In both of these solutions the~influence of the~non-local gravitational effects of the~bulk on the~brane are represented by a~single \uvozovky{braneworld} parameter $b$. The~$1/r^2$ behaviour of the~second term in the~Newtonian potential

\begin{equation}
 \Phi=-\frac{M}{M^2_\mathrm{P}r}+\frac{b}{2r^2}\label{beq9}
\end{equation}
inspired the~name \uvozovky{tidal charge} for the~parameter b \cite{Dad-etal:2002:PHYLB:}.
 \par

\section{Orbital motion in the~braneworld Kerr spacetimes}

\begin{figure}[t]
\begin{center}
\includegraphics[width=0.9\linewidth]{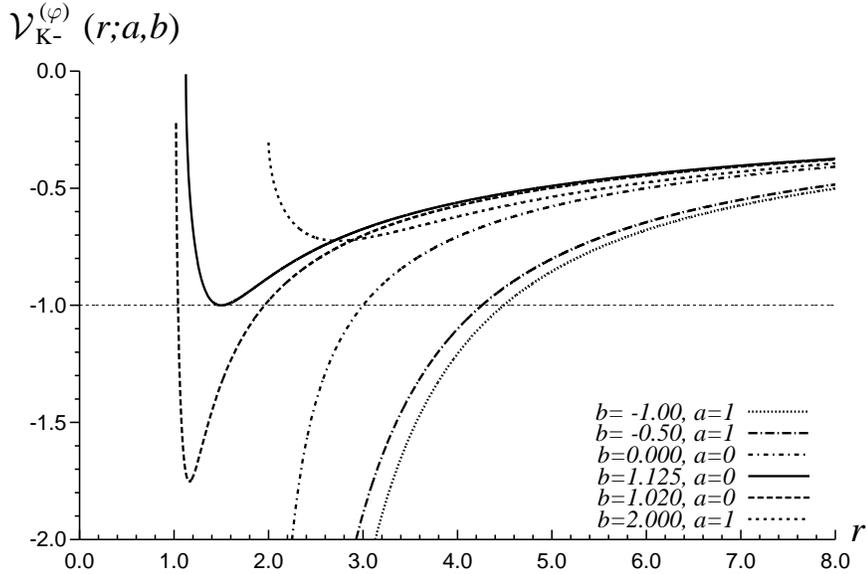}
\caption{\label{Vmin} Minus-family orbital velocity profiles in braneworld Kerr black hole and naked singularity spacetimes.} 
\end{center}
\end{figure}

\begin{figure}[ht]
\begin{center}
\begin{minipage}{.49\linewidth}
\centering
\includegraphics[width=\linewidth]{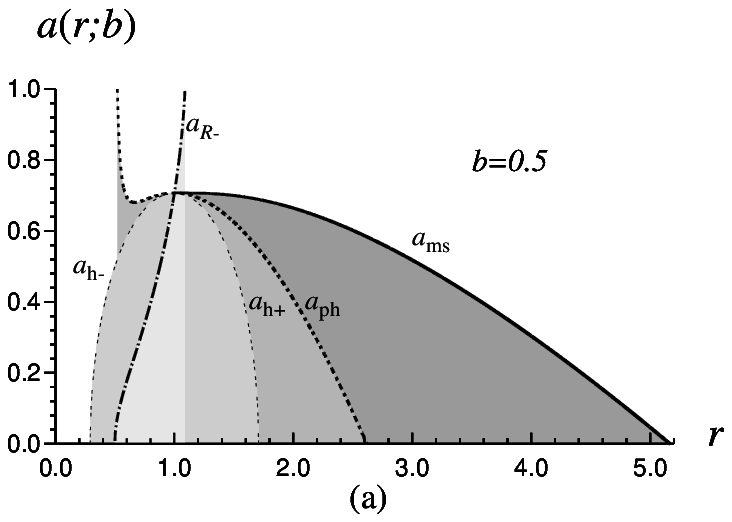}
\end{minipage}\hfill
\begin{minipage}{.49\linewidth}
\centering
\includegraphics[width=\linewidth]{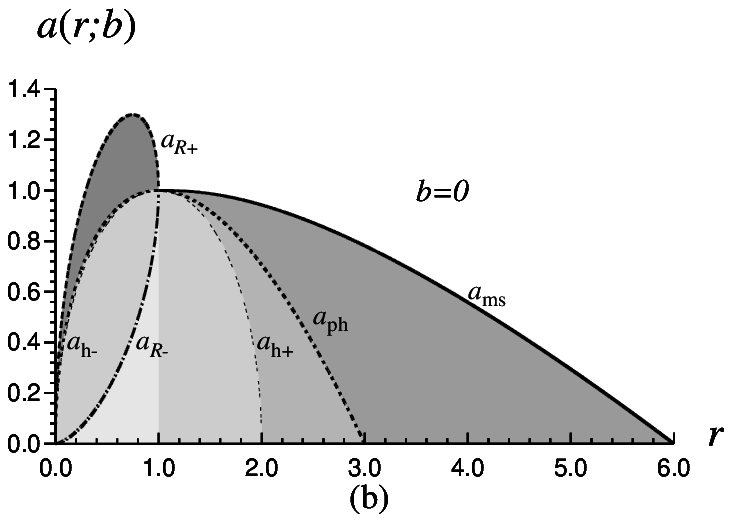}
\end{minipage}
\begin{minipage}{.49\linewidth}
\centering
\includegraphics[width=\linewidth]{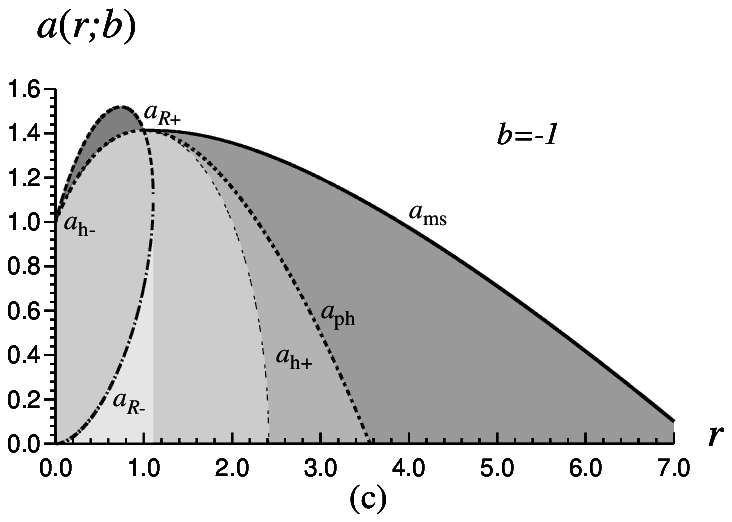}
\end{minipage}\hfill
\begin{minipage}{.49\linewidth}
\centering
\includegraphics[width=\linewidth]{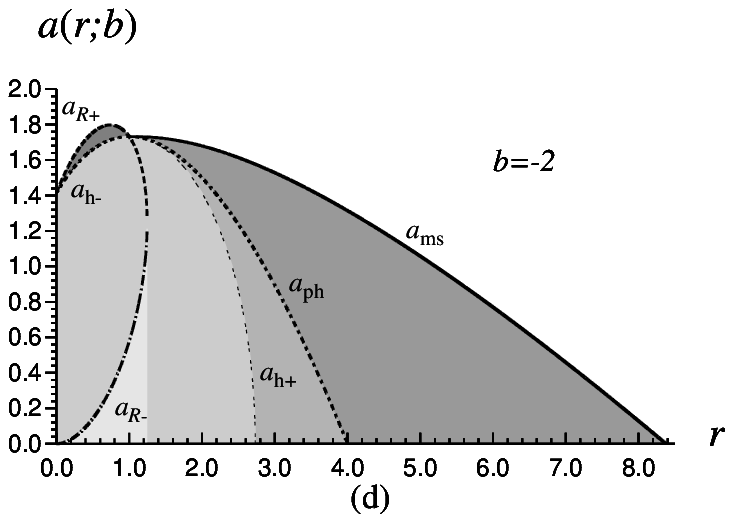}
\end{minipage}\hfill
\caption{\label{aphams} The~loci of the~photon circular orbits $a_{\mathrm{ph}}(r,b)$; the~marginally stable orbits $a_{\mathrm{ms}}(r,b)$; function $a_{R\pm}(r,b)$ and function $a_{\mathrm{h\pm}}(r,b)$ defined implicitly by position of event horizon for appropriately chosen fixed values of $b$. }
\end{center}
\end{figure}

\begin{figure}[ht]
\begin{center}
\includegraphics[width=\linewidth]{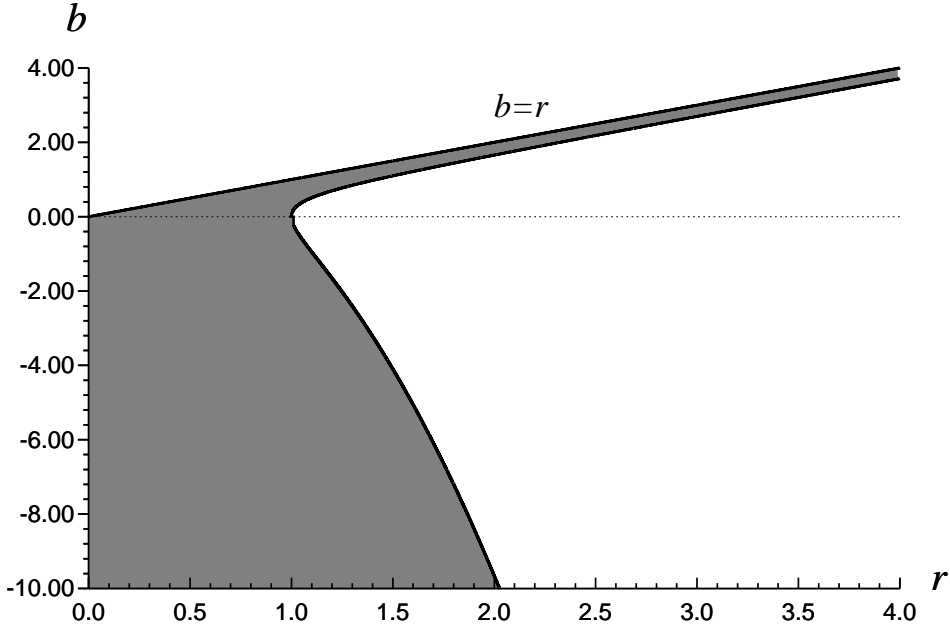}
\caption{\label{arad} Naked singularity regions with potentially retrograde plus-family orbits.}
\end{center}
\end{figure}

\begin{figure}[t]
\begin{center}
\includegraphics[width=\linewidth]{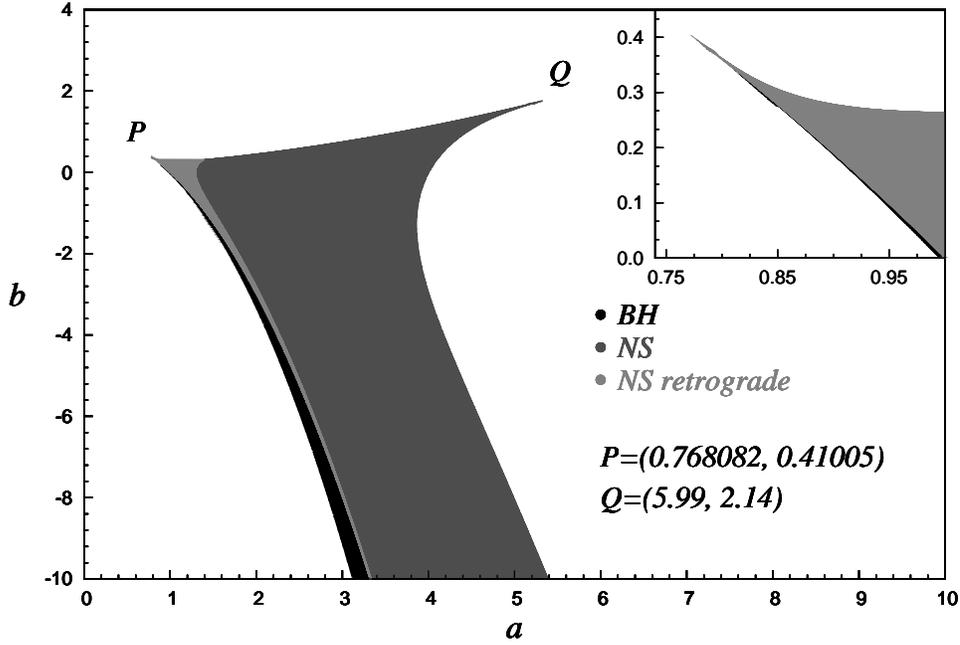}\\
\caption{\label{Oregion}Classification of the~braneworld Kerr spacetimes according to existence of the~Aschenbach effect. The~Aschenbach effect is allowed in the~black region representing black holes,  dark-grey region representing naked singularities with corotating orbits only, and lighter-grey region representing naked singularities with retrograde motion in the~LNRF-velocity profile (corresponding to negative values of the~function $\mathcal{V}^{(\phi)}_{\mathrm{K}}$). }
\end{center}
\end{figure}

\begin{figure}[t]
\begin{center}
\includegraphics[width=0.9\linewidth]{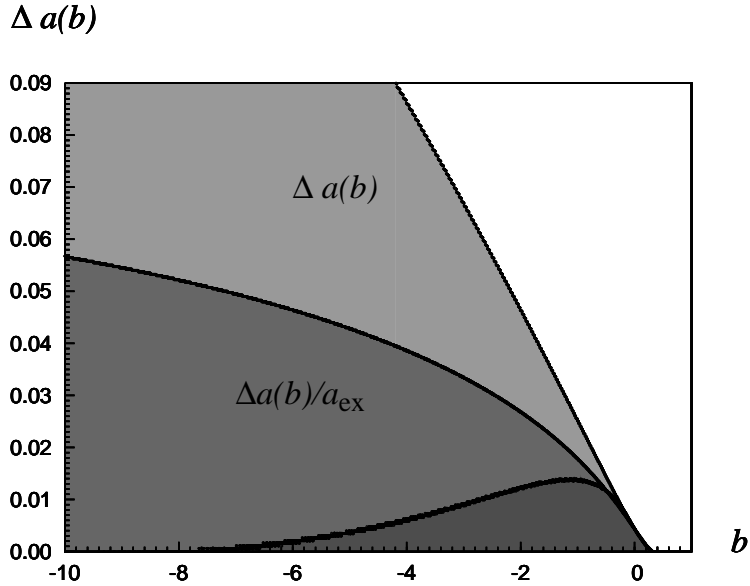}
\caption{\label{Oregion2}Spin range of braneworld Kerr black holes allowing for the~Aschenbach effect given as a~function of the~tidal charge $b$. It is illustrated by the~function $\Delta a(b)$ and the~relative spin range determined by the~function $\Delta a(b)/a_\mathrm{ex}$. The~darkest area represents most interesting case (from astrophysical point of view) when the~radius of marginally stable orbit $r_{\mathrm{ms}}$ is less than the~local minimum of the~Keplerian velocity profile.}
\end{center}
\end{figure}

\begin{figure}[ht]
\begin{minipage}{.49\linewidth}
\centering
\includegraphics[width=\linewidth]{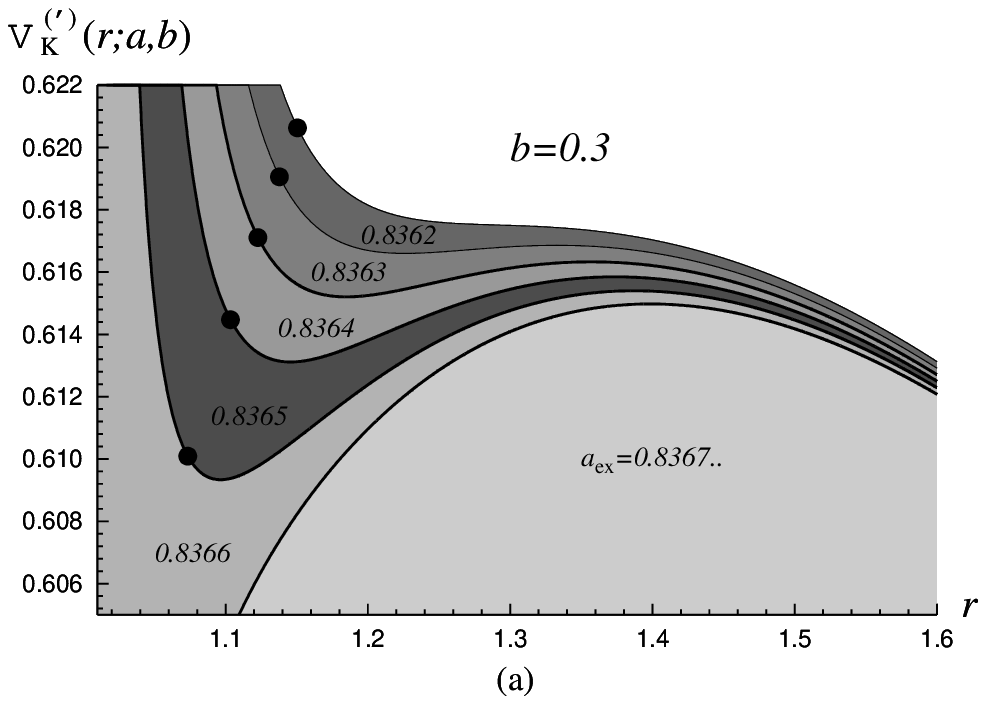}
\end{minipage}\hfill
\begin{minipage}{.49\linewidth}
\centering
\includegraphics[width=\linewidth]{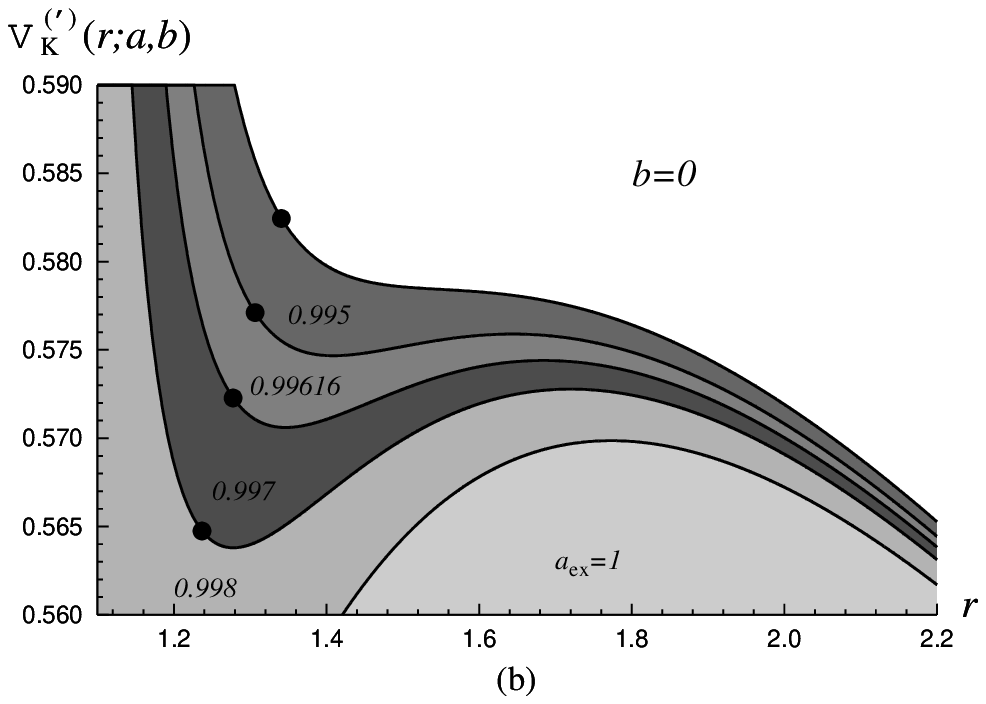}
\end{minipage}\hfill
\begin{minipage}{.49\linewidth}
\centering
\includegraphics[width=\linewidth]{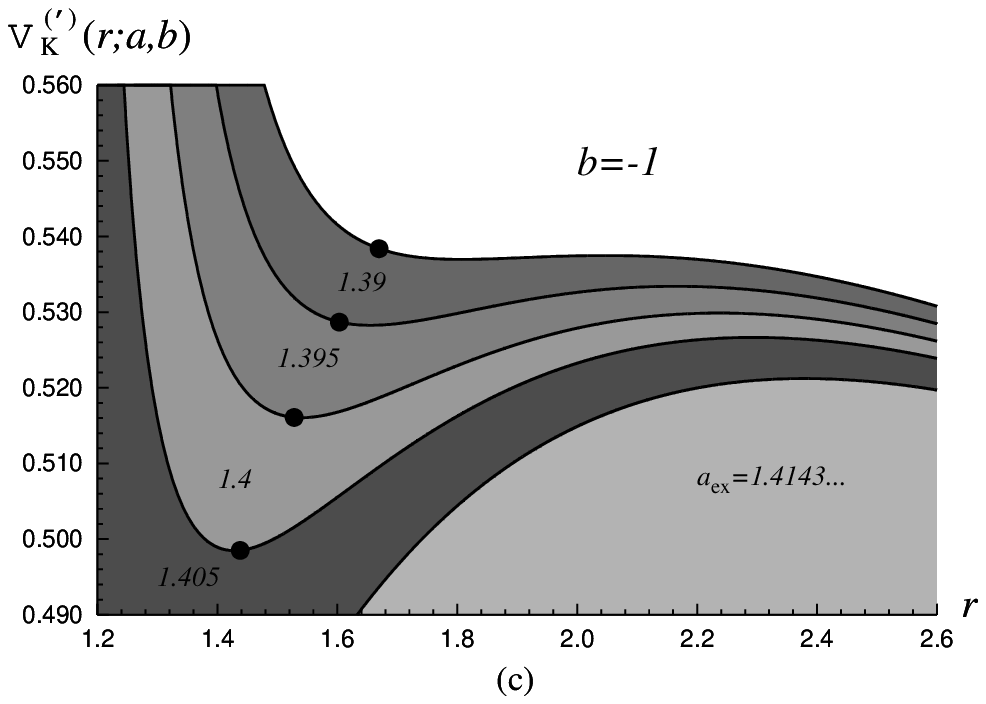}
\end{minipage}\hfill
\begin{minipage}{.49\linewidth}
\centering
\includegraphics[width=\linewidth]{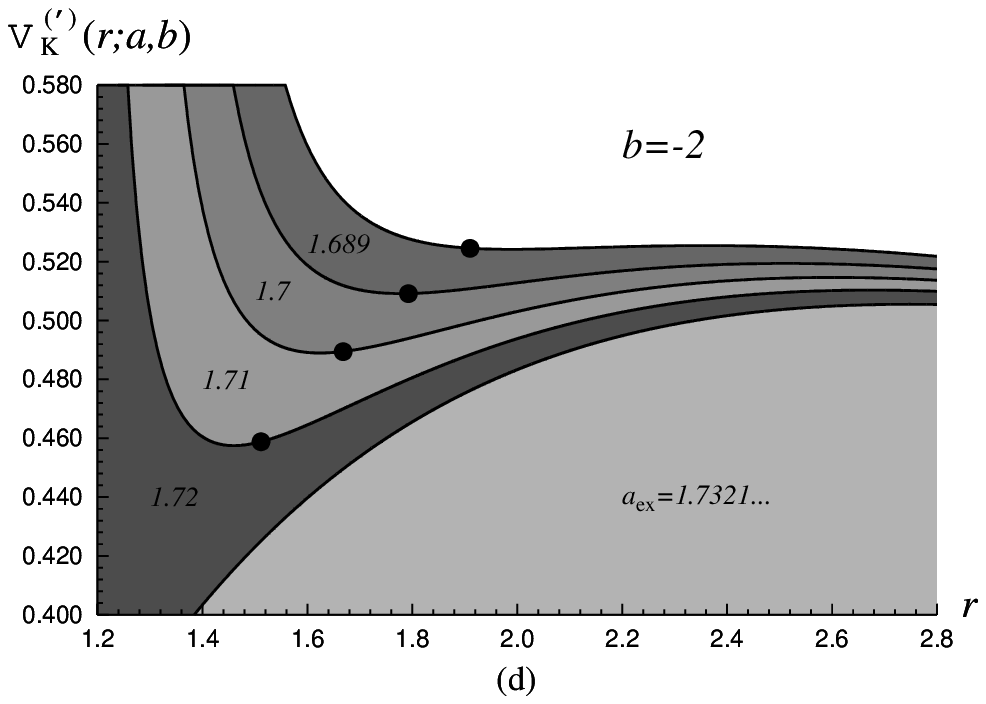}
\end{minipage}\hfill
\caption{\label{Ovk} Non-monotonic LNRF-related velocity profiles for braneworld Kerr black hole backgrounds given for some values of the~tidal charge $b$ and appropriately chosen values of the~$a$. The black points denote loci of $r_\mathrm{ms}$.}
\end{figure}

Motion of test particles in the~field of braneworld rotating black holes is given by the~geodesic structure of the~Kerr--Newman spacetimes with the~tidal charge $b$. The~braneworld parameter reflects the~tidal effects of the~bulk space and has no influence on the~motion of charged particles. The~geodesic structure given by the~Carter equations \cite{Car:1973:BlaHol:} is relevant for both uncharged and charged test particles. The~circular test particle orbits of the~braneworld Kerr black holes are identical to the~circular geodesics of the~Kerr--Newman spacetime with properly chosen charge parameter. 

We shall study the~Aschenbach effect, i.e., we look for the~non-monotonicity (humps) in the~LNRF-velocity profiles of Keplerian discs orbiting near-extreme braneworld Kerr black holes or naked singularities. 


\subsection{Geometry}
Using standard Boyer-Lindquist coordinates $(t,r,\theta,\varphi)$ and geometric units $(c=G=1)$, we can write the~line element of rotating (Kerr) black hole on the~3D-brane in the~form 
\[
\mathrm{d}s^2= - \left(1-\frac{2Mr - b}{\Sigma }\right)\mathrm{d}t^2 - \frac{2a(2Mr - b)}{\Sigma}\,\mathrm{sin}^2\theta 
\,\mathrm{d}t\,\mathrm{d}\phi\ + 
\]
\begin{equation}\label{Metrika}
+ \frac{\Sigma}{\Delta}\,\mathrm{d}r^2 + \Sigma\, \mathrm{d}\theta^2 + \left(r^2 + a^2 + \frac{2Mr - b}
{\Sigma}\,a^2\mathrm{sin}^2\theta\right)
\mathrm{sin}^2\theta\, \mathrm{d}\phi^2\, ,
\end{equation}
where 
\begin{eqnarray}
\Delta = r^2 -2Mr +a^2 + b\, ,\\ 
\Sigma = r^2 + a^2\mathrm{cos}^2\theta\, ,
\end{eqnarray}
$M$ and $a=J/M$ are the~mass parameter and the~specific angular momentum of the~background, while the~braneworld parameter $b$, called \uvozovky{tidal charge}, represents the~imprint of non-local (tidal) gravitational effects of the~bulk space \cite{Ali-Gum:2005:}. The~physical \uvozovky{ring} singularity of the~braneworld rotating black holes (and naked singularities) is located at $r=0$ and $\theta = \pi/2$, as in the~Kerr spacetimes.

The~form of the~metric (\ref{Metrika}) is the~same as that of the~standard Kerr--Newman solution of the~4D Einstein-Maxwell equations, with tidal charge $b$ being replaced by squared electric charge $Q^2$ \cite{Mis-Tho-Whe:1973:Gra:}. The~stress tensor on the~brane $\mathcal{E}_{\mu\nu}$ takes the~form

\begin{eqnarray}
\mathcal{E}^t_t &=& -\mathcal{E}^\varphi_{\varphi} = -\frac{b}{\Sigma^3}\left[\Sigma - 2(r^2 + a^2)\right]\, ,\\
\mathcal{E}^r_r &=& - \mathcal{E}^\theta_\theta = -\frac{b}{\Sigma^2}\, ,\\
\mathcal{E}^t_\varphi &=& - (r^2 + a^2)\,\mathrm{sin}^2 \mathcal{E}^\varphi_t = -\frac{2ab}{\Sigma^3}\left(r^2 + a^2\right)\mathrm{sin}^2\theta\, ,
\end{eqnarray} 
that is fully analogical $(b\rightarrow Q^2)$ to components of the~electromagnetic energy-momentum tensor of the~Kerr--Newmann solution in Einstein's general relativity \cite{Ali-Gum:2005:}. For negative values of the~tidal charge ($b<0$), the~values of the~black hole spin $a>M$ are allowed. Such a~situation is forbidden for the~standard 4D Kerr black holes. In the~following, we put $M=1$ in order to work with completely dimensionless formulae.

\subsection{Locally non-rotating frames and orbital motion}

The~orbital velocity of matter orbiting a~braneworld Kerr black hole along circular orbits is given by appropriate projections of its 4-velocity $U=(U^t,0,0,U^\varphi)$ onto the~tetrad of a~locally non-rotating frame (LNRF) \cite{Bar-Pre-Teu:1972:ASTRJ2:} 
\begin{equation}\label{LNRF}
\mathrm{\bf e}^{(t)} = \left ( \omega^2g_{\phi\phi} - g_{tt}\right ) ^{\frac{1}{2}}{\bf d}t\ ,
\end{equation}
\begin{equation}
\mathrm{\bf e}^{(\phi)} = (g_{\phi\phi})^{\frac{1}{2}}({\bf d}\phi - \omega{\bf d}t)\, ,
\end{equation}
\begin{equation}
\mathrm{\bf e}^{(r)} = \left(\frac{\Sigma }{\Delta }\right)^{\frac{1}{2}}{\bf d}r\ ,
\end{equation}
\begin{equation}
\mathrm{\bf e}^{(\theta)} = \Sigma^{\frac{1}{2}}{\bf d}\theta\ ,
\end{equation}
where $\omega$ is the~angular velocity of the~LNRF relative to distant observers and reads
\begin{equation}\label{omega}
\omega =-\frac{g_{t\phi}}{g_{\phi\phi}}= \frac{a(2r-b)}{\Sigma(r^2 + a^2) + (2r - b)\,a^2\,\mathrm{sin^2\theta}}\, .
\end{equation}

For the~circular motion, the~only non-zero component of the~3-velocity measured locally in the~LNRF is the~azimuthal  component that is given by

\begin{eqnarray}\label{V}
\mathcal{V}^{(\phi)}_{\mathrm{LNRF}} &=\frac{[\Omega - \omega]}{\sqrt{\left(\omega^2- \frac{g_{tt}}{g_{\phi\phi}} \right)}}\nonumber\\ &= \nonumber
\frac{\left[\left(r^2 + a^2\right)^2-a^2\Delta\,\mathrm{sin}^2\,\theta\,\right]\mathrm{sin}\,\theta\,(\Omega-\omega)}{\Sigma\sqrt{\Delta}}\, ,
\end{eqnarray}
where 
\begin{equation}
\Omega = \frac{U^{\phi}}{U^{t}}= -\frac{lg_{tt}+g_{t\phi}}{lg_{t\phi} + g_{\phi\phi}}
\end{equation}
is the~angular velocity of the~orbiting matter relative to distant observers and 
\begin{equation}
l=-\frac{U_\phi}{U_{t}}
\end{equation} 
is its specific angular momentum; $U_t\, , U_\phi$ are the~covariant components of the~4-velocity field of the~orbiting matter. 

Using (\ref{Metrika}) we arrive to the~formula
\begin{equation}\label{Vomega}
\Omega = \frac{\left(1-\frac{2r-b}{\Sigma}\right)l + \frac{a(2r-b)}{\Sigma}\,\mathrm{sin}^2\theta}
{\left(r^2 + a^2 + \frac{2r -b}{\Sigma}\,a^2\,\mathrm{sin}^2\theta\right)\mathrm{sin}^2\theta
-l\frac{a(2r-b)}{\Sigma}\,\mathrm{sin}^2\theta}\, .
\end{equation}

\section{Velocity profiles for the~Keplerian distribution of the~specific angular momentum}

\subsection{Circular geodesics}

Motion of test particles following circular geodetical orbits in the~equatorial plane $(\theta = \pi/2)$ is described by the~Keplerian distribution of the~specific angular momentum, which in the~braneworld Kerr backgrounds takes the~form \cite{Stu-Kot:2009:GRG:}
\begin{equation}\label{lk}
l_{\mathrm{K\pm }}(r;a,b) = \pm \frac{(r^2 + a^2)\sqrt{r-b}\mp a(2r-b)}{r^2 -2r +b \pm a\sqrt{r-b}}\, ; 
\end{equation}
the~signs $\pm $ refer to two distinct families of orbits in the~Kerr braneworld spacetimes. From the~LNRF point of view the~2nd family, or minus-family (given by the~lower sign), represents retrograde orbits, while the~1st family, or plus-family (given by the~upper sign), represents direct orbits in the~black hole spacetimes \cite{Bar-Pre-Teu:1972:ASTRJ2:}, but can represent both direct and retrograde orbits in the~naked singularity spacetimes. We can find a~similiar situation for Kerr spacetime in \cite{Stu:1980:BULAI:,Stu:1981:BULAI:}. For both families, we find a~formal limit of the~Keplerian motion to be located at $r = b$. In the~black hole spacetimes, it is located under the~inner horizon and is irrelevant from the~astrophysical point of view. In the~naked singularity spacetimes, it is relevant for $b>0$, while for $b<0$, it is located under the~ring singularity at $r=0$ which represents the~limit on the~location of circular orbits.

The~corresponding Keplerian angular velocity related to distant observers is given by the~relation
\begin{equation}\label{Vomegak}
\Omega_\mathrm{K\pm }(r;a,b) =\pm\frac{1}{r^2/\sqrt{r-b}\pm a}\, ,
\end{equation}
and the~Keplerian orbital velocity related to the~LNRF in braneworld Kerr backgrounds is thus given by the~relation 
\begin{equation}\label{Vk}
\mathcal{V}^{(\phi)}_{\mathrm{K\pm }}(r;a,b) =\pm \frac{\sqrt{r-b}(r^2+a^2)\mp  a(2r- b)}
{\left(r^2\pm  a\sqrt{r-b}\right)\sqrt{\Delta}}\ .
\end{equation}
Clearly, there is a~limit on existence of circular equatorial geodesics at $r=b$ that is relevant (and located above the~ring singularity) for spacetimes with positive tidal charge. 

Like in the~Kerr backgrounds $(b=0)$, also in the~braneworld Kerr backgrounds only the~plus-family orbits exhibit the~\uvozovky{minimum-maximum} humpy structure of the~LNRF-related orbital velocity profiles $\mathcal{V}^{(\phi)}_{\mathrm{K}}(r)\,$. On the~other hand, in braneworld Kerr naked-singularity spacetimes with positive tidal charge $(b>0)$, there is always at least one local extreme in the~orbital velocity profiles (maximum in $\mathcal{V}^{(\varphi)}_{\mathrm{K+}}$ and minimum in $\mathcal{V}^{(\varphi)}_{\mathrm{K-}}$ profiles) where the~orbital velocity gradient changes its sign. This non-monotonicity, however, does not correspond to the~Aschenbach effect. 


We illustrate the~typical character of $\mathcal{V}^{(\phi)}_{\mathrm{K+}}(r;a,b)$ velocity profiles in the~braneworld rotating black hole and naked singularity spacetimes with negative tidal charge in Figure \ref{Ovkns} -- in this case the~profiles are extended down to $r=0$, but here we do not consider region of $r<0$. For special values of parameters of naked singularity spacetimes, the~Aschenbach effect can be related to retrograde region of 1st family orbits (the humpy structure contains orbits with $\mathcal{V}^{(\varphi)}_{K+}<0$).

For positive tidal charge the~characteristic profiles of $\mathcal{V}^{(\phi)}_{\mathrm{K+}}$ are presented in the~Figure \ref{Vkod} -- in this case the~profiles finish their validity at $r=b$. There is also depicted velocity profile under the~inner horizon of the~black hole. Numerical calculations indicate that for those kinds of velocity profiles there is no Aschenbach effect, so for the~black hole spacetimes we shall focus our attention on the~behaviour of the~Keplerian profiles $\mathcal{V}^{(\phi)}_{\mathrm{K+}}(r;a,b)$ above the~outer horizon. 

For both positive and negative tidal chrges the~profiles of $\mathcal{V}^{(\phi)}_{\mathrm{K-}}$ in black hole and naked singularity spacetimes are presented in Figure \ref{Vmin}. These profiles do not exhibit any \uvozovky{minimum-maximum} structures. Therefore, in the following we shall focus attention to the~plus-family orbits only and use notation $\mathcal{V}^{(\phi)}_{\mathrm{K}}(r;a,b)$ instead of $\mathcal{V}^{(\phi)}_{\mathrm{K+}}(r;a,b)$.

We have to consider the~function $\mathcal{V}^{(\phi)}_\mathrm{K}(r;a,b)$ (and also the~functions $l_\mathrm{K}(r;a,b)$, $\Omega_\mathrm{K}(r;a,b)$) within the~range of definition of the~Keplerian motion \cite{Stu-Kot:2009:GRG:}. The~range is governed by the~radii of the~photon circular geodesics $r_{\mathrm{ph}}$ given implicitly by the~relation

\beq
           a = a_{\mathrm{ph}}(r;b) \equiv \frac{r(3-r)-2b}{2\sqrt{r-b}}\, ,
\eeq
and by the~radii of the~marginally stable circular geodesics $r_{\mathrm{ms}}$, implicitly given by 

\beq
           a = a_{\mathrm{ms}}(r;b) \equiv \frac{4(r-b)^{3/2}\mp r\sqrt{3r^2-2r(1+2b)+3b}}{3r-4b}\, .
\eeq
The~functions $a_{\mathrm{ph}}(r;b)$ and $a_{\mathrm{ms}}(r;b)$ are illustrated in Figure \ref{aphams} -- for detailed discussion of the~properties of photon and marginally stable orbits see \cite{Stu-Kot:2009:GRG:,Kot-Stu-Tor:2008:CLAQG:}. Above the~black hole outer horizon, the~stable orbits are located in the~interval of $r_{\mathrm{ms}} < r < \infty$, while unstable orbits are located in the~interval $r_{\mathrm{ph}} < r < r_{\mathrm{ms}}$. Note that for Kerr naked singularity the~situation is generally more complicated, see for example \cite{Stu-Kot:2009:GRG:}. For $r \rightarrow r_{\mathrm{ph}}$ there is $\mathcal{V}^{(\phi)}_\mathrm{K}(r;a,b) \rightarrow 1$. The~stable circular orbits are relevant for Keplerian accretion discs, therefore, it is reasonable to put the~limits on the~physical relevance of the~Aschenbach effect to the~region of stable circular geodesics. 


\begin{figure}[p]
\begin{minipage}{.45\linewidth}
\centering
\includegraphics[width=\linewidth]{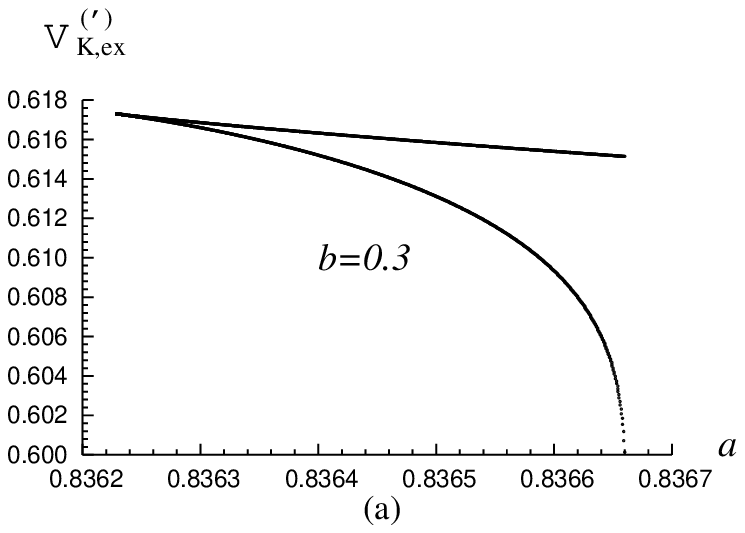}
\end{minipage}\hfill
\begin{minipage}{.45\linewidth}
\centering
\includegraphics[width=\linewidth]{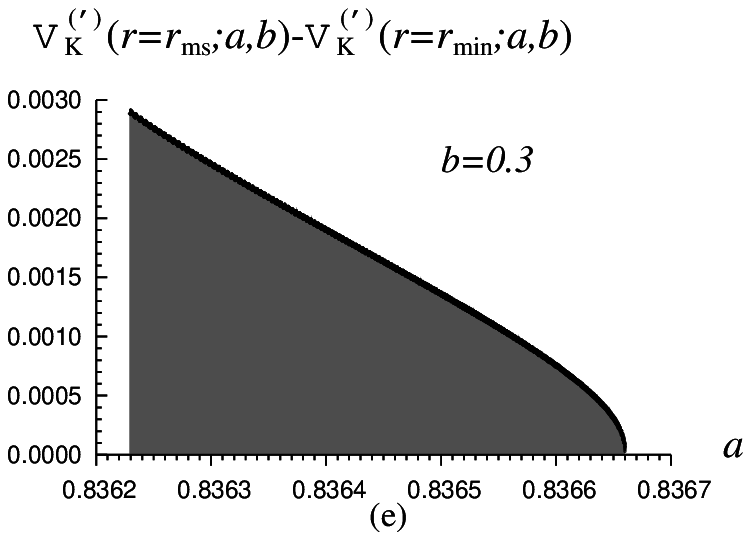}
\end{minipage}\hfill
\begin{minipage}{.45\linewidth}
\centering
\includegraphics[width=\linewidth]{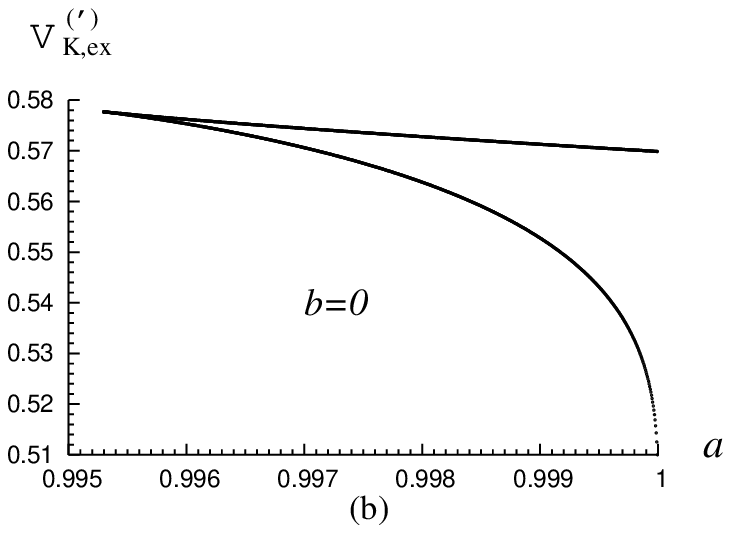}
\end{minipage}\hfill
\begin{minipage}{.45\linewidth}
\centering
\includegraphics[width=\linewidth]{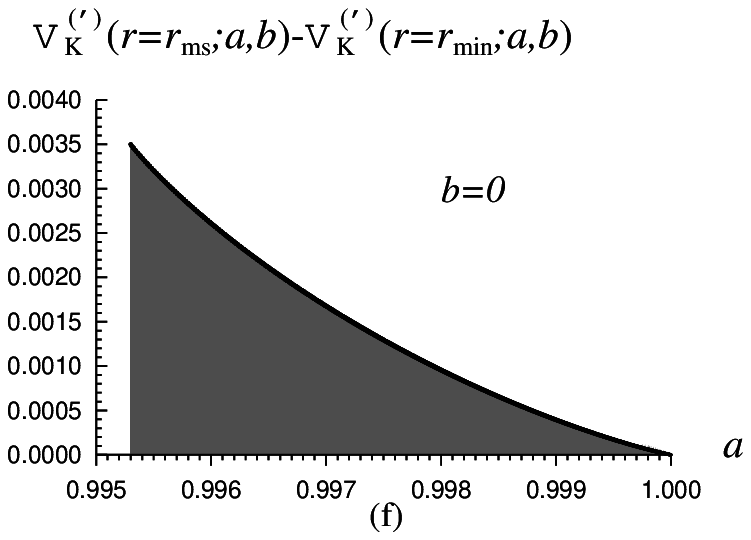}
\end{minipage}\hfill
\begin{minipage}{.45\linewidth}
\centering
\includegraphics[width=\linewidth]{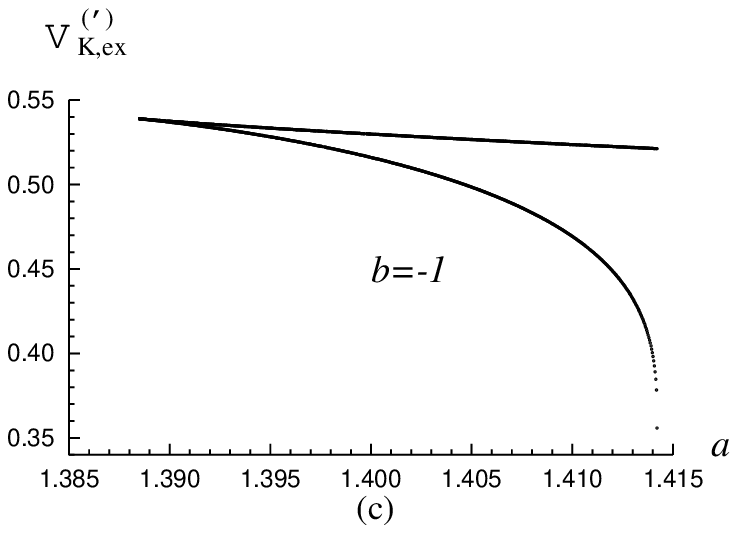}
\end{minipage}\hfill
\begin{minipage}{.45\linewidth}
\centering
\includegraphics[width=\linewidth]{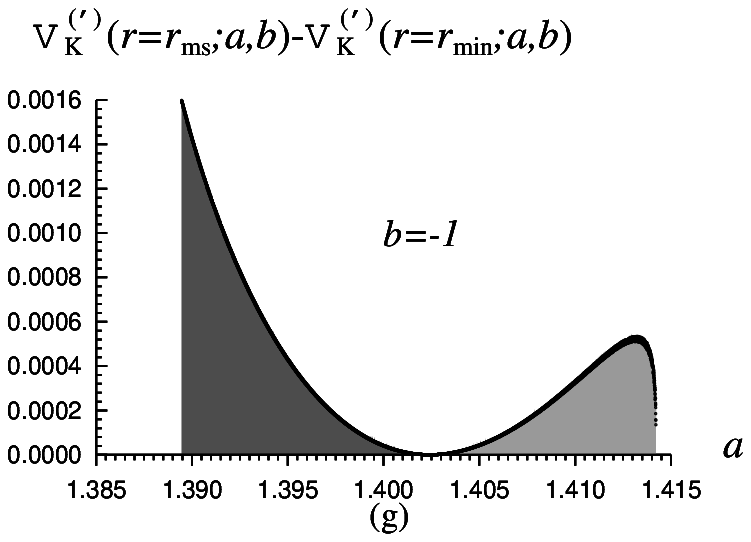}
\end{minipage}\hfill
\begin{minipage}{.45\linewidth}
\centering
\includegraphics[width=\linewidth]{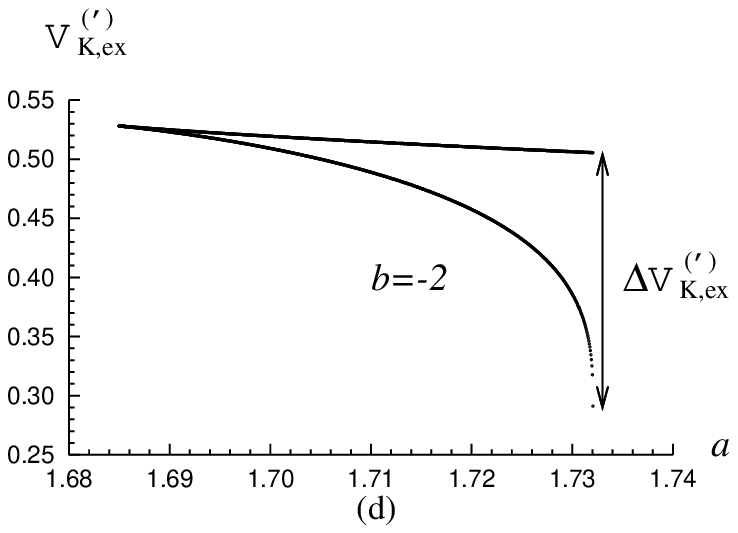}
\end{minipage}\hfill
\begin{minipage}{.45\linewidth}
\centering
\includegraphics[width=\linewidth]{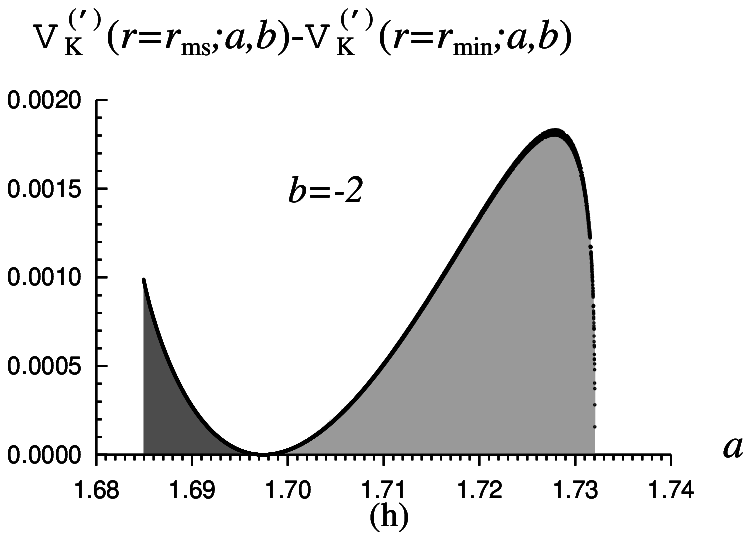}
\end{minipage}\hfill
\caption{\label{Ovex} 

(a)--(d) Function $\mathcal{V}^{(\phi)}_{\mathrm{K,ex}}$ which determines values of the~local minimum and maximum of the~function $\mathcal{V}^{(\phi)}_{\mathrm{K}}$. The~upper line represents the~local maximum, while the~lower line represents the~local minimum. These curves demonstrate that maximum of the~velocity difference is reached for the~extreme black hole states. 
In (d), there is an indication for the~definition of extremal velocity difference (occurring for the~extreme black hole states) as a~function of the~tidal charge $b$, function $\Delta\mathcal{V}^{(\phi)}_{\mathrm{K,ex}}$ giving maximal difference between value of the~local maximum and local minimum of the~function $\mathcal{V}^{(\phi)}_{\mathrm{K,ex}}$, considered with fixed $b$, is in Figure \ref{Ovex2}. 

(e)--(h) Function $\mathcal{V}^{(\phi)}_{\mathrm{K}}(r=r_{\mathrm{ms}},a,b)-\mathcal{V}^{(\phi)}_{\mathrm{K}}(r=r_{\mathrm{min},a,b})$ which defines difference between $\mathcal{V}^{(\phi)}_{\mathrm{K}}$ with radius for marginally stable orbit and with radius for local minimum. The~dark grey region represent more astrophysicaly interesting cases, when radius of marginally stable orbit $r_{\mathrm{ms}}$ is lower than the~radius for the~local minimum $r_{\mathrm{min}}$, see Figure \ref{Oregion2}} 
\end{figure}

It is well known that for standard 4D Kerr black holes the~Aschenbach effect, i.e., the~non-monotonic LNRF-related orbital velocity profile appears for strongly limited class of near-extreme black holes \cite{Asch:2004:ASTRA:,Stu-Sla-Tor-Abr:2005:PRD:}. It appears in the~regions where the~LNRF-related velocity of Keplerian motion reaches relatively large magnitude $\mathcal{V}^{(\phi)}_\mathrm{K} \sim 0.5-0.6$, but the~velocity difference of the~minimum-maximum hump is much smaller ($\Delta \mathcal{V}^{(\phi)}_\mathrm{K} \sim 0.01$). These are the~reasons why the~effect was overlooked for a~relatively long time. On the~other hand, the~Aschenbach effect is much stronger for Kerr naked singularities and is manifested for a~large range of spin $1 < a < 4.0005$. Moreover, for Kerr naked singularities with spin close to the~extreme black hole state ($a=1$), the~Aschenbach effect is connected to another interesting effect related to circular geodesics -- namely the~retrograde character of the~1st family circular geodesics related to the~LNRF. The~counter-rotating orbits of the~1st family can constitute a~part of the~non-monotonic Keplerian profile (see Figure \ref{class} and Figure \ref{Ovkns}). The~\uvozovky{retrograde} Kerr naked singularities manifesting (strongly) the~Aschenbach effect, can be determined by the~relation (implied by the~condition $\mathcal{V}^{(\phi)}_\mathrm{K} = 0$)

\beq
        a = a_{R\pm}(r) \equiv \sqrt{r}\left(1 \pm \sqrt{1-r}\right)\, .
\eeq
It is illustrated in Figure \ref{aphams} (the~case $b=0$) -- we see that the~retrograde 1st-family orbits exist at radii $0<r<1$, for spin parameters $1 < a_{R+} < 3\sqrt{3}/4$; the~Keplerian LNRF-related velocity profile touches $\mathcal{V}^{(\phi)}_\mathrm{K} = 0$ at $r=1$ for $a = 1$. 

For braneworld Kerr naked singularities, the~retrograde motion of plus-family circular geodesics appears for spin determined by the~condition $a_{R-} < a < a_{R+}$, where 

\beq
         a_{R\pm}(r;b) \equiv \frac{(2r-b) \pm \sqrt{\mathcal{D}}}{2\sqrt{r-b}}\, ,
\eeq
where
\beq
         \mathcal{D} = b^2-4r(1-r)b +4r^2(1-r)\, .
\eeq
Functions $a_{R\pm}(r;b)$ are illustrated in Figure \ref{aphams}.
The~conditions $\mathcal{D} \geq 0$ and $r>b$ put limit on the~radii where for given tidal charge $b$ the retrograde plus-family orbits can exist, as illustrated in Figure {\ref{arad}}. 

\subsection{Aschenbach effect in braneworld spacetimes} 

Character of the~LNRF-related velocity profile of the~Keplerian (equatorial) circular motion is determined by the~behaviour of the~velocity gradient that can be expressed in the~form

\begin{equation}\label{vkder}
\frac{\partial \mathcal{V}^{(\phi)}_\mathrm{K}}{\partial r}=\frac{A_1}{A_2}-\frac{A_3A_4}{A_2^2}\, ,
\end{equation}
where
\begin{equation}
A_1= \sqrt{r-b}\left(\frac{r^2+a^2}{2(r-b)}+2r\right) -2a\, ,
\end{equation}
\begin{equation}
A_2= \left(a\sqrt{r-b}+ r^2\right)\sqrt{\Delta}\, ,
\end{equation}
\begin{equation}
A_3=\sqrt{r-b}\left(r^2+a^2\right) -a(2r-b)\, ,
\end{equation}
\begin{equation}
A_4=\sqrt{\Delta}\left[\left(\frac{a}{2\sqrt{r-b}}+2r\right)+\frac{r-1}{\Delta}\left(a\sqrt{r-b}+r^2\right)\right]\, .
\end{equation}

Considering the~braneworld Kerr black holes, we restrict our attention to the~region above the~event horizon at $r > r_{+} = 1 + \sqrt{1-a^2-b}$. The~rotation (spin) parameter of  black hole spacetimes is limited by 

\beq
      a_{\mathrm{ex}}=\sqrt{1-b}\, 
\eeq
for a~given tidal charge $b$. When braneworld Kerr naked singularities are considered, the~region $r>b$ has to be studied for the~existence of the~humpy LNRF-velocity profiles when $b>0$, while for $b<0$ we have to analyse the~whole region of $r>0$ above the~ring singularity. 

Local extrema of $\mathcal{V}^{(\phi)}_\mathrm{K}(r)$ profiles (giving their \uvozovky{minimum-maximum} humpy parts) are determined by the~condition $\partial \mathcal{V}^{(\phi)}_\mathrm{K}/\partial r = 0\,$, i.e., by zero points of the~function (\ref{vkder}) that are identical to the~roots of the~polynomial 
\begin{equation}\label{root}
g(Z):\sqrt{r}\left(h_1 Z^4 + h_2 Z^3 + h_3 Z^2 + h_4 Z + h_5\right) =0\ ,
\end{equation}
where
\begin{eqnarray}
h_1&=&r\left[3r^2 + 2(1-2b)r-3b\right]\, ,\\
h_2&=&-2r^2\left(1-b/r\right)^{3/2}(3r+1)\, ,\\
h_3&=&4r^4-2(5+3b)r^3+16br^2+2br(1-3b)-b^2\, ,\\
\nonumber h_4&=&\sqrt{1-b/r}\left\{2r^3\left[4\left(1-b/r\right)^2+4\left(1-b/r\right)+1\right]\right. - \\
&-&\left. 2r^4\left[1+4\left(1-b/r\right)\right] \right\}\, ,\\
h_5&=&r^3(r^2-2br+b)\, .
\end{eqnarray}

\begin{figure}[ht]
\includegraphics[width=\linewidth]{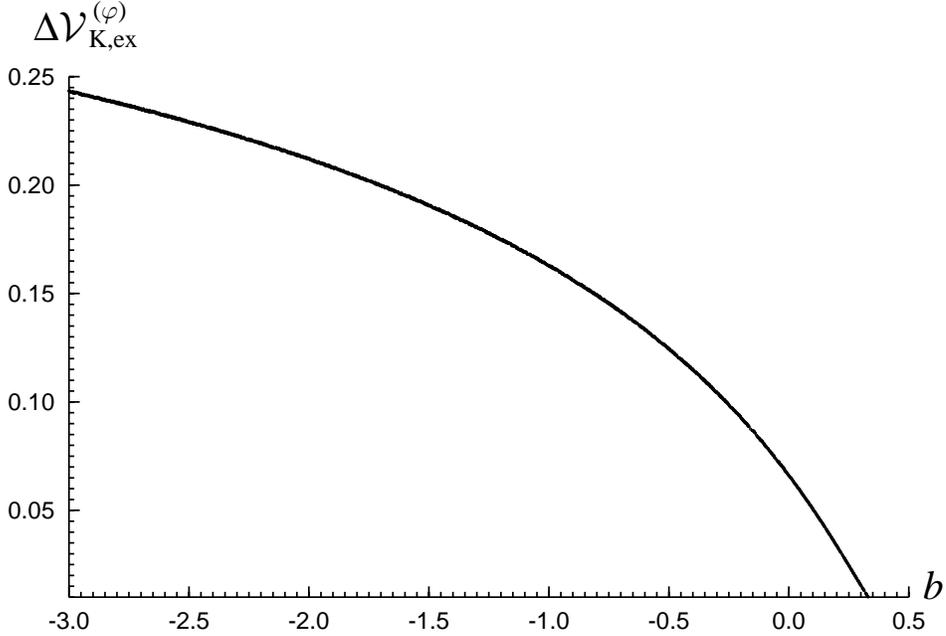}
\caption{\label{Ovex2} Function $\Delta\mathcal{V}^{(\phi)}_{\mathrm{K,ex}}$ giving maximal difference between values of the~local maximum and local minimum of the~function $\mathcal{V}^{(\phi)}_{\mathrm{K,ex}}$, for a~fixed $b$.} 
\end{figure}

\begin{figure}
\begin{minipage}{.49\linewidth}
\centering
\includegraphics[width=\linewidth]{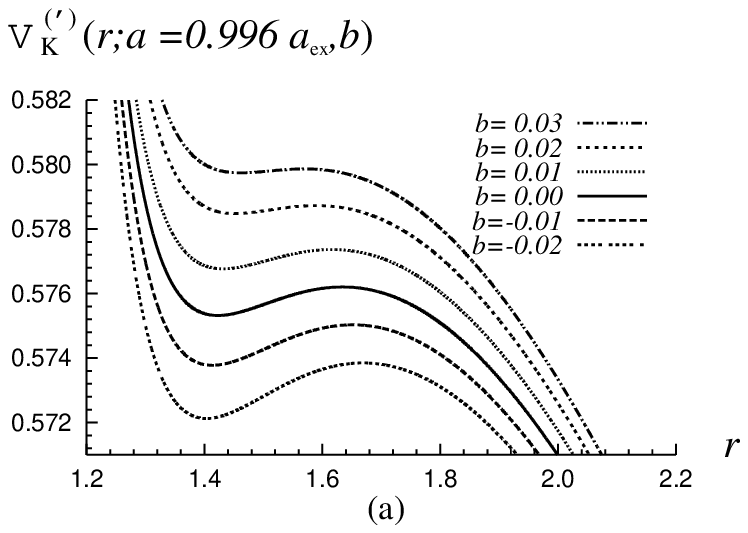}
\end{minipage}\hfill
\begin{minipage}{.49\linewidth}
\centering
\includegraphics[width=\linewidth]{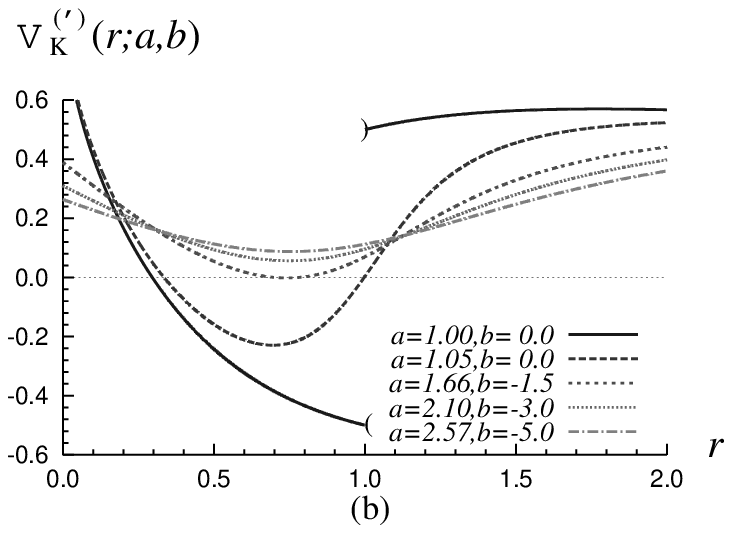}
\end{minipage}\hfill
\caption{\label{Ovkns2} 
(a) Sequence of non-monotonic LNRF-related velocity profiles for near extreme black hole background. In the~sequence, there is $a=0.996\, a_\mathrm{ex}$ and we change the~parameter $b$. 

(b) Sequence of non-monotonic LNRF-related velocity profiles for naked singularity background close to the~extreme black hole state given by $a_\mathrm{ex} = \sqrt{1-b}$. In the~sequence, there is $a=1.05\, a_\mathrm{ex}$, and we can see the~transform between the~retrograde and purely corotating velocity profiles. For comparison, the~velocity profile constructed for the~extreme Kerr black hole is included.}
\end{figure}

The~occurrence of the~Aschenbach effect can be analysed by numerical study of the~roots of Eq. (\ref{root}). The~numerical approach enables to determine the~region in the~parameter space ($a - b$) where this effect is admitted, see Figure \ref{Oregion}.  In naked singularity region, there is a~subregion with retrograde 1st-family orbits.
We can see that for black holes the~region is highly sensitive on the~choice of the~tidal charge $b$. For positive values of $b$ increasing, it becomes narrowed as compared with the~case of $b=0$ and disappears when $b>0.41005$. In the~limit value of $b=0.41005$, the~corresponding value of the~spin reads $a=0.76808$. On the~other hand, the~interval of the~black hole spin allowing for the~Aschenbach effect enlarges significantly with descending of negative tidal charge $b$.  We demonstrate this dependence in Figure \ref{Oregion2} presenting the~function $\Delta a(b)$, which denotes spin interval of the~black holes (with tidal charge $b$) allowing for the~Aschenbach effect. Since the~maximal black hole spin $a_{\mathrm{ex}}$ also depends on $b$, we give the~dependence of the~ratio $\Delta a/ a_{\mathrm{ex}}$ on $b$ for completeness. Notice that for tidal charge $b=-1$, the~spin interval $\Delta a$ (and even its relative magnitude given by $\Delta a/ a_{\mathrm{ex}}$) increases by almost one order as compared to the~case of $b=0$. In the~naked singularity spacetimes, the~Aschenbach effect appears in wide region of $a,b$ parameters -- both for positive and negative tidal charges. The~allowed naked singularity region is limited by $b=2.14$ and $a=5.99$ -- the~point $Q$ in Figure \ref{Oregion2}. 

The~behaviour of humpy LNRF-velocity profiles of Keplerian orbits in the~field of braneworld Kerr black holes is  represented by a~series of figures for both positive and negative tidal charges. In order to clearly illustrate all the~aspects of the~Aschenbach effect in dependence on the~black hole parameters, we give figures with both fixed value of the~tidal charge $b$ and fixed value of the~rotation parameter $a$, see Figure \ref{Ovk}. (The~boundary of stable orbits given by the~marginally stable orbit is represented by a~point on all the~profiles.) Moreover, we include the~dependences of the~LNRF velocities in the~minimum and maximum of the~humpy profile on the~black hole parameters (see Figure \ref{Ovex}). These extremal LNRF-related velocities decrease with the~tidal charge descending, but their difference increases significantly -- for fixed $b$, the~depth of the~humpy profile grows with black holes spin approaching $a_\mathrm{ex} = \sqrt{1-b}\,$. Finally, in Figure \ref{Ovex2}, there is the~dependence of the~velocity difference at the~minimum-maximum part of the~humpy profile on the~tidal charge $b$ for extreme braneworld Kerr black holes. We can se that for $b=-1$ the~extremal difference increases by a~factor $\sim 2$ as compared to the~case of $b=0$. 
 
We also demonstrate there, how the~Aschenbach effect disappears about the~point $Q$ for positive tidal charges, and how the~LNRF-related Keplerian profiles in the~naked singularity spacetime can be reduced into those for the~corresponding extreme black hole spacetime, when the~retrograde naked singularity profile is transformed into a~discontinuity occurring at the~radius $r=1$, and the~negatively valued part under the~horizon becomes physically irrelevant. Behaviour of Keplerian velocity profiles in both black hole and naked singularity backgrounds corresponding to near-extreme black hole cases is shown in the~Figure \ref{Ovkns2}.

\section{Conclusions}

We have shown that the~Aschenbach effect is a~typical feature of the~circular geodetical motion in the~field of both standard and braneworld Kerr naked singularities with a relatively large interval of spins above the~extreme black-hole limit. For naked-singularity spin sufficiently close to the~extreme black-hole state, the~Aschenbach effect is manifested by the~retrograde plus-family circular orbits. For black hole spacetimes, such retrograde orbits can appear under the~inner horizon, being thus irrelevant from the~astrophysical point of view. In the~field of near-extreme rotating black holes, the~Aschenbach effect located above the~outer black hole horizon can be thus considered as a~small remnant of typical naked singularity phenomenon. 

We have demonstrated that in the~braneworld near-extreme Kerr black hole spacetimes, the~non-monotonic LNRF-related orbital velocity profiles of the~Keplerian motion are suppressed for positive tidal charges as compared with the~standard Kerr black holes, and disappear for the~tidal charge $b > 0.41005$, while they are strongly enlarged for decreasing negative tidal charge. The~spin range of black holes allowing for the~Aschenbach effect increases significantly with decreasing tidal charge, strengthening possible relevance of the~Aschenbach effect in astrophysical phenomena (see, e.g., \cite{Svo-Dov-Goo-Kar:2009:ASTRA:}). This possibility is further supported by increasing magnitude of the~velocity difference in the~minimum-maximum part of the~LNRF-related Keplerian profile with decreasing negative tidal charge. Recent observations of high-frequency quasiperiodic oscillations observed in the~X-ray spectra of some neutron-star binaries \cite{Tor-Abr-Bak-Bur-Klu-Reb-Stu:2008:actaa:,Stu-Kot-Tor:2008:Actaa:,Tor:2009:AAP:,Bou-Barr-Lin-Tor:2010:} that could be used to test braneworld models in the~strong-field limits imply negatively-valued dimensionless tidal charges as high as $b=-2$ \cite{Kot-Stu-Tor:2008:CLAQG:} when the~Aschenbach effect could be quite significant and well observed \cite{Svo-Dov-Goo-Kar:2009:ASTRA:}. 

We conclude that the~Aschenbach effect can play a~significant role in explaining a~variety of physical phenomena (optical effects and related line profiles, explanation of a~variety of high-frequency quasi-periodic oscillations observed in some microquasars and active galactic nuclei, etc.) expected to appear in the~strong gravitational field of braneworld Kerr black holes, especially in the~case of negative tidal charge.

\section*{Acknowledgment}
The~presented work was supported by the~Czech grants MSM~4781305903, LC 06014, GA{\v C}R 205/09/H033 and the~internal grant SGS/2/2010. 

\section*{References}

\end{document}